\documentclass[lettersize,journal]{IEEEtran}
\IEEEoverridecommandlockouts
\usepackage{cite}
\usepackage{amsmath,amssymb,amsfonts}
\usepackage{algorithmic}
\usepackage{graphicx}
\usepackage{textcomp}
\usepackage{xcolor}
\def\BibTeX{{\rm B\kern-.05em{\sc i\kern-.025em b}\kern-.08em
    T\kern-.1667em\lower.7ex\hbox{E}\kern-.125emX}}
\usepackage{multirow}
\usepackage{algorithm}
\usepackage{array}
\usepackage[caption=false,font=normalsize,labelfont=sf,textfont=sf]{subfig}
\usepackage{textcomp}
\usepackage{stfloats}
\usepackage{url}
\usepackage{xcolor}
\usepackage{array}
\usepackage{verbatim}
\usepackage{caption} 
\usepackage{graphicx}
\usepackage{subfig}
\usepackage{enumerate}
\usepackage{enumitem}
\usepackage{tabularx,booktabs,textcomp}
\usepackage{amsthm}

\usepackage{amsmath,amssymb,amsthm}

\usepackage{enumitem}
\makeatletter 
\newcommand{\linebreakand}{%
  \end{@IEEEauthorhalign}
  \hfill\mbox{}\par
  \mbox{}\hfill\begin{@IEEEauthorhalign}
}
\makeatother

\begin{document}

\title{Fault Classification in Electrical Distribution Systems using
Grassmann Manifold}

\author{{K. Victor Sam Moses Babu,~\IEEEmembership{Member,~IEEE}, Sidharthenee Nayak,  Divyanshi Dwivedi,~\IEEEmembership{Student Member,~IEEE}, Pratyush Chakraborty,~\IEEEmembership{Senior Member,~IEEE}, Chandrashekhar Narayan Bhende,~\IEEEmembership{Senior Member,~IEEE}, Pradeep Kumar Yemula,~\IEEEmembership{Member,~IEEE}, Mayukha Pal,~\IEEEmembership{Senior Member,~IEEE}}

\thanks{(Corresponding author: Mayukha Pal)}

\thanks{Mr. K. Victor Sam Moses Babu is a Data Science Research Intern at ABB Ability Innovation Center, Hyderabad 500084, India and also a Research Scholar at the Department of Electrical and Electronics Engineering, BITS Pilani Hyderabad Campus, Hyderabad 500078, IN.}
\thanks{Ms. Sidharthenee Nayak is a Data Science Research Intern at ABB Ability Innovation Center, Hyderabad 500084, India, and also a post graduate student at the School of Electrical Sciences, Indian Institute of Technology, Bhubaneswar, 751013, IN}
\thanks{Ms. Divyanshi Dwivedi is a Data Science Research Intern at ABB Ability Innovation Center, Hyderabad 500084, India, and also a Research Scholar at the Department of Electrical Engineering, Indian Institute of Technology, Hyderabad 502205, IN.}
\thanks{Dr. Pratyush Chakraborty is an Asst. Professor with the Department of Electrical and Electronics Engineering, BITS Pilani Hyderabad Campus, Hyderabad 500078, IN.}
\thanks{Dr. Chandrashekhar Narayan Bhende is a Professor with the School of Electrical Sciences
Indian Institute of Technology
Bhubaneswar, 751013, IN.}
\thanks{Dr. Pradeep Kumar Yemula is an Assoc. Professor with the Department of Electrical Engineering, Indian Institute of Technology, Hyderabad 502205, IN.}
\thanks{Dr. Mayukha Pal is with ABB Ability Innovation Center, Hyderabad-500084, IN, working as Global R\&D Leader – Cloud \& Analytics (e-mail: mayukha.pal@in.abb.com).}
}

\maketitle

\begin{abstract}

Electrical fault classification is vital for ensuring the reliability and safety of power systems. Accurate and efficient fault classification methods are essential for timely and effective maintenance. In this paper, we propose a novel approach for effective fault classification through Grassmann manifolds, which is a non-Euclidean space that captures the intrinsic structure of high-dimensional data and offers a robust framework for feature extraction. We use simulated data for electrical distribution systems with various types of electrical faults. The proposed method involves transforming the measurement fault data into Grassmann manifold space using techniques from differential geometry. This transformation aids in uncovering the underlying fault patterns and reducing the computational complexity of subsequent classification steps. To achieve fault classification, we employ a machine learning technique optimized for the Grassmann manifold. The support vector machine classifier is adapted to operate within the Grassmann manifold space, enabling effective discrimination between different fault classes. The results illustrate the efficacy of the proposed Grassmann manifold-based approach for electrical fault classification which showcases its ability to accurately differentiate between various fault types.
\end{abstract}


\section{Introduction}

Electrical distribution systems (EDS) are susceptible to various types of faults which include ground and short circuit faults. Ground faults occur when an unintended path between an electrical current and a grounded element is established due to insulation breakdown. Short circuit faults happen when a low-resistance connection forms between two points in an electrical system which leads to excessive current flow. These faults could cause significant disruptions in the power distribution system, leading to equipment damage, power outages, and potential safety hazards \cite{intro1}. 

The increasing demand for electrical power supply has led to the integration of distributed generators (DGs) in distribution systems which offers several advantages, including a reduction in environmental pollution, efficient power generation, decreased losses in transmission lines, improved voltage profiles, and a high-quality power supply to consumers \cite{intro2}. However, there are certain challenges in terms of changing fault current levels and affecting the coordination with protective switchgear. The fault current levels vary depending on the type of distributed energy resources (DERs) connected to the system, leading to the mal-operation of conventional relays designed for fault detection and classification. Photovoltaic (PV) introduces power quality issues due to nonlinear power electronics devices and loads. The integration of electric vehicles and bidirectional electric vehicle charging stations in distribution networks further increases operational requirements, posing additional challenges to the safe and reliable operation of the distribution networks \cite{RAI2021106914}.

To address these challenges, accurate and fast fault classification is essential and digital relays play a crucial role in achieving this objective which relies on precise and rapid algorithms. Performing fault classification in EDS is crucial and indispensable for swiftly identifying and localizing faults within the system \cite{intro2}. Fault classification helps operators and maintenance personnel efficiently isolate faulty components while expediting the restoration of power to unaffected areas. The coordination of protective devices which include relays relies on the accurate classification of faults, ensuring that these devices respond appropriately to safeguard the integrity of the broader electrical network. Also, fault classification aids in implementing safety measures during maintenance which reduces risks to personnel and equipment. By optimizing protective devices based on fault types, utilities enhance the overall reliability and effectiveness of the protection system. Additionally, fault classification data facilitates preventive maintenance planning which enables utilities to proactively address potential issues before they escalate into faults. The analysis of fault patterns supports continuous system improvement to contribute to a resilient, safe, and reliable power distribution infrastructure while minimizing downtime \cite{PRASAD201848}.

In recent years the advancements in wavelet techniques \cite{wavelet1,wavelet2,wavelet3,wavelet4,wavelet5}, machine learning (ML) \cite{ml1,ml2,ml3,RAI2021106914,VAISH2021104504}, deep learning (DL) based techniques \cite{dl1,dl2,dl3,dl4,dl5,dl6}, and fuzzy logic \cite{fuzzy1,fuzzy2,fuzzy3,fuzzy4} approaches have provided the way for the development of various fault detection and classification algorithms in power system networks. The wavelet transform is a robust signal-processing technique for fault classification in EDS \cite{wavelet1}. Through its ability to analyze signals concurrently in both time and frequency domains; excels in detecting transient events and capturing high-frequency components associated with faults. The process typically involves acquiring voltage or current signals, preprocessing to remove noise, and applying wavelet transform \cite{wavelet2}. Following this, the relevant features are extracted from the wavelet coefficients which include energy distribution and statistical measures across different scales and time intervals. Feature selection helps in reducing computational complexity, leading to the application of machine learning algorithms or neural networks for fault identification. However, considerations such as the choice of wavelet basis, optimal scale selection, and real-time implementation challenges must be taken into account for successful application in fault classification \cite{wavelet3}.

 ML techniques such as support vector machines (SVM), random forests, and decision trees are employed to extract relevant features from electrical signals and classify faults based on patterns identified during training on labeled datasets \cite{ml1,ml2}. Feature engineering and selection enhance the efficiency of these models by making them adept at recognizing distinct fault types. On the other hand, the DL model uses artificial neural networks (ANNs) with multiple hidden layers to capture intricate patterns and hierarchical representations in signal data. Convolutional neural networks (CNNs) excel in spatial pattern recognition, while recurrent neural networks (RNNs) are effective for modeling sequential aspects in time-series signals \cite{dl3,dl4}. The application of autoencoders for unsupervised feature learning and transfer learning for using pre-trained models further enhances fault classification capabilities \cite{auto}. Challenges include the need for high-quality labeled data and considerations of interpretability in scenarios where human understanding of the classification process is essential. However, ML/ DL techniques \cite{DD_pmu,Victor_EVCI,DD_resilience} contribute to automated and accurate fault detection in electrical distribution systems by offering solutions to the complexities of the data and specific application requirements. These advanced techniques use the power of data analysis and pattern recognition to improve the overall reliability and efficiency of power system networks by ensuring a prompt response to faults, minimizing interruptions to the network, and improving the resilience of the system \cite{GNN, Victor_IAS}.

In this paper, the autoregressive moving average (ARMA) Grassmann method for fault classification in EDS is proposed. It is an innovative approach that uses the Grassmann manifold and integrates it with an ARMA  model and a support vector machine (SVM) classifier \cite{Al-Samhi_2021}. The Grassmann manifold is a mathematical concept that represents the space of all k-dimensional subspaces of an n-dimensional vector space. It has a geometric structure and finds applications in areas such as computer vision, signal processing, and machine learning \cite{LI2024105491}. The proposed methodology adopts a comprehensive 10-fold cross-validation strategy, ensuring robust model performance assessment and generalization capabilities. The dataset undergoes feature extraction within each fold by projecting it onto the Grassmann manifold which captures linear subspaces. This representation transforms the data for subsequent analysis. The SVM model equipped with a Gaussian kernel helps in handling non-linear relationships which is trained on the Grassmann manifold features using the training set and then evaluated on the test set. Performance metrics such as accuracy, precision, recall, and F1 score are employed to validate the model's efficacy. The manifold representation combined with the non-linear capabilities of the SVM Gaussian kernel offers a unique perspective for fault classification, especially in handling high-dimensional signal data from electrical distribution systems. This innovative approach offers a novel perspective by treating fault data as points on a Grassmann manifold. Leveraging the manifold's geometric properties enhances the effectiveness of classification. This unique methodology could address the complexities and variabilities in fault data more effectively, resulting in improved accuracy and robustness in EDS fault classification. In the study by Rai et al. \cite{RAI2021106914}, the authors conducted a thorough comparison of various neural network models, achieving high accuracy. However, the proposed model stands out for its computational efficiency, ease of implementation, and reduced memory requirements, all while maintaining accuracy equal to or surpassing that of traditional neural network models.

The key contributions of the work are as follows:

\begin{enumerate}
    \item An innovative framework for capturing the intrinsic structures and variations within fault data using Grassmann manifold as a representation space for electrical signals.
    \item The integration of the ARMA model adds a temporal dynamics component to the fault classification methodology. This inclusion is useful for capturing the time-dependent characteristics of fault signals, enhancing the model's ability to discern patterns and anomalies in the electrical distribution system over time.
    \item The use of an SVM with a Gaussian kernel allows the model to handle non-linear relationships in the data. This contribution is crucial for effectively capturing complex fault patterns and relationships that may exist within electrical signals.
\end{enumerate}

The paper is structured as follows: Section \ref{section:Methods} provides a detailed exploration of the ARMA model for temporal data representation and investigates the Grassmann manifold. The adaptation of SVM kernels for effective use on the Grassmann manifold is discussed. Section \ref{section:Simulation} outlines the parameters and initial conditions of the simulation study, with a discussion of obtained results. Finally, Section \ref{section:Conclusion} summarizes the paper.

\section{Materials and Methods}
\label{section:Methods}

\begin{figure}[b]
  \centering
  \includegraphics[width=3.5in]{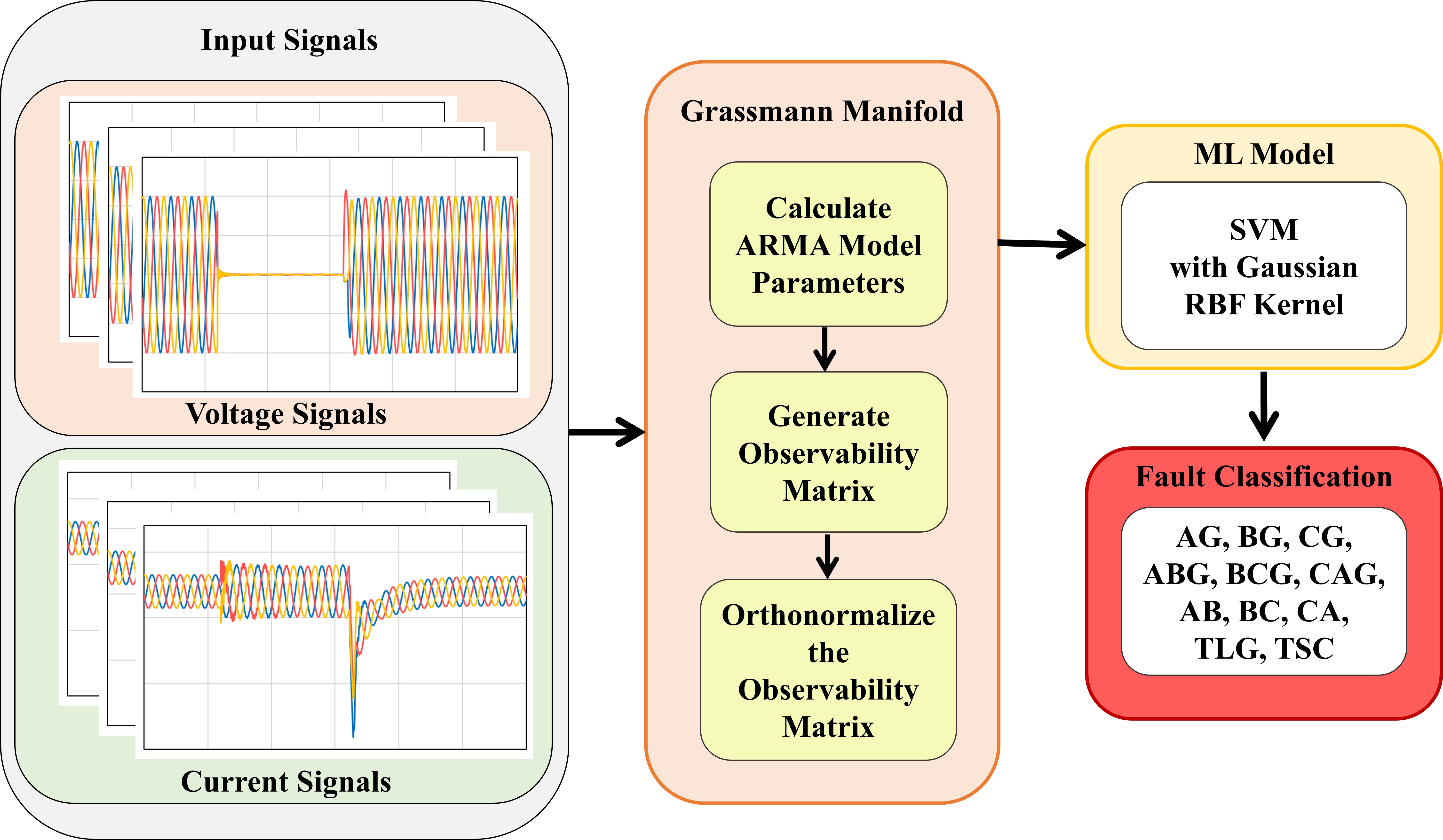}
  \caption{Proposed Methodology.}
  \label{fig:process}
\end{figure}

The architecture of our proposed model for signal classification is illustrated in Fig. \ref{fig:process}. The model operates in three main stages: autoregressive moving average processing, Grassmann manifold feature extraction, and machine learning classification. 

\subsection{ARMA model}
Spatio-temporal datasets are frequently analyzed using dynamic models, as highlighted in various studies \cite{turaga2011statistical, al2021time}. One common approach is employing ARMA (Autoregressive Moving Average) models, which are essential in time series analysis. An ARMA model integrates two distinct models: the Autoregressive (AR) model, which predicts future data points based on past observations, and the Moving Average (MA) model, which utilizes historical forecast errors for prediction.

Consider a time-series dataset denoted as \emph{x}. The current value of \emph{x} may depend on its own previous values and on the past values of another series \emph{u} and is represented as,
\begin{equation}
\sum_{j=0}^{r}P(j) ; x(t-j) = \sum_{j=0}^{s}Q(j) ; u(t-j)
\end{equation}
where, \emph{x(t)} signifies the observed outputs, and $P(j), Q(j)\nabla \in \mathbb{R}^{5\times5}$ are the parameter matrices for the AR and MA coefficients, respectively. The unobserved inputs are modeled as white noise:
\begin{equation}
E\mathcal{N}(j)=0, E\mathcal{N}(u)\eta(t)=\delta_{ut}\Theta
\end{equation}

\begin{equation}
p(z) = \sum_{j=0}^{r}P(j)z^{j}
\end{equation}
\begin{equation}
q(z) = \sum_{j=0}^{s}Q(j)z^{j}
\end{equation}

The ARMA model \cite{turaga2011statistical} can be described as:
\begin{equation}
g(t)=Hx(t)+k(t) , k(t)\sim N(0,S)
\end{equation}
\begin{equation}
x(t+1)=Px(t)+m(t) , m(t)\sim N(0,T)
\end{equation}
where, $x \in \mathbb{R}^{d}$ is the hidden state vector, $P \in \mathbb{R}^{d\times d}$ is the state transition matrix, and $H \in \mathbb{R}^{r\times d}$ is the observation matrix. The observed features are denoted as $g \in \mathbb{R}^{r}$, while $k$ and $m$ represent noise components, assumed to be normally distributed with zero mean and covariances $S \in \mathbb{R}^{r\times r}$ and $T \in \mathbb{R}^{d\times d}$, respectively.

In dealing with high-dimensional time-series data, such as video or dynamic textures, a common method is to first determine a lower-dimensional representation of the observations through principal component analysis (PCA), and then learn the temporal dynamics in this reduced space. Suppose the observations $g(1), g(2),..., g(\tau)$ represent the features at time indices $1, 2, ..., \tau$. Let $[g(1), g(2),..., g(\tau)] = L\Lambda M^{T}$ be the singular value decomposition of the dataset. The model then becomes:
\begin{equation}
\hat{H}=L , \hat{P}=\Lambda M^{T} E_{1} M (M^{T} E_{2} M)^{-1}\Lambda^{-1}
\end{equation}

\begin{equation}
E_{1} =
\begin{bmatrix}
0 & 0 \ I_{\tau -1} & 0
\end{bmatrix} ,
E_{2} =
\begin{bmatrix}
I_{\tau -1} & 0 \ 0 & 0
\end{bmatrix}
\end{equation}

For the ARMA model in equation (6), starting from an initial state $x(0)$, the expected sequence of observations is given by \cite{5206710}:
\begin{equation}
\mathbb{E}
\begin{bmatrix}
g(0) \, g(1) \, g(2) \dots
\end{bmatrix}
=
\begin{bmatrix}
H \, HP \, HP^{2} \dots
\end{bmatrix}
x(0)=
O_{\infty}(N) x(0)
\end{equation}

In practical applications, the infinite observability matrix is approximated by a finite observability matrix \cite{990925}:
\begin{equation}
O_{l}^{T}=[ H^{T} (HP)^{T} (HP^{2})^{T} \dots (HP^{l-1})^{T}]
\end{equation}

\subsection{Grassmann Manifolds}
Grassmann manifolds, which represent linear subspaces, are crucial in a variety of applications, including machine learning, computer vision, image processing, low-rank matrix optimization, dynamic low-rank decompositions, and model reduction. The Grassmann manifold, denoted as \emph{$G_{l}(d,r)$}, is the set of all linear subspaces of a fixed dimension 
$r$ within the Euclidean space $\mathbb{R}^{d}$. This set is also known as the Grassmannian.

In mathematical terms, the Stiefel Manifold, represented as $S_{v}(d,r)$, comprises all ordered orthonormal $r-tuples$ in $\mathbb{R}^{d}$:
\begin{equation}
S_{v}(d,r) = {L\in \mathbb{R}^{{d} \times {r}} | L^{T}L=I_{r} }
\end{equation}

To provide manifold structures to $G_{l}(d,r)$ and $S_{v}(d,r)$, we view these sets of matrices as quotients of the orthogonal group:
\begin{equation}
O(d) = {R\in \mathbb{R}^{{d} \times {d}} | R^{T}R=I_{d}=RR^{T} }
\end{equation}
Thus, the Grassmann manifold is defined by constructing the equivalence class of all orthogonal matrices under the orthogonal group $O(d)$:
\begin{equation}
G_{l}(d,r) = {S_{v}(d,r)/O(d)}
\end{equation}

As per \cite{al2021time}, the expected outcomes $g(t)$ are found within the column space of the observability matrix $O_{\infty}$, and $O_{\infty}$ is approximated to form $O_{l} \in M_{lr \times d}$ by truncating at the 
$l^{th}$ block. The ARMA model thus represents a signal as a Euclidean subspace (defined by the column space of $O_{l}$), corresponding to a point on the Grassmann manifold $G_l (d,lr)$. Let $[X_{1}]$ and $[X_{2}]$ denote two points on the Grassmann manifold. The distance between these two points differs from the standard Euclidean distance. The projection distance between two points on the Grassmann manifold \cite{Jayasumana2014KernelMO} is given by:
\begin{equation}
d_{j}([X_{1}],[X_{2}]) = 2^{-0.5}||X_{1} , X_{1}^{T} - X_{2}, X_{2}^{T}||_{F}
\end{equation}

\begin{figure*}[b]
  \centering
  \includegraphics[width=7.0in]{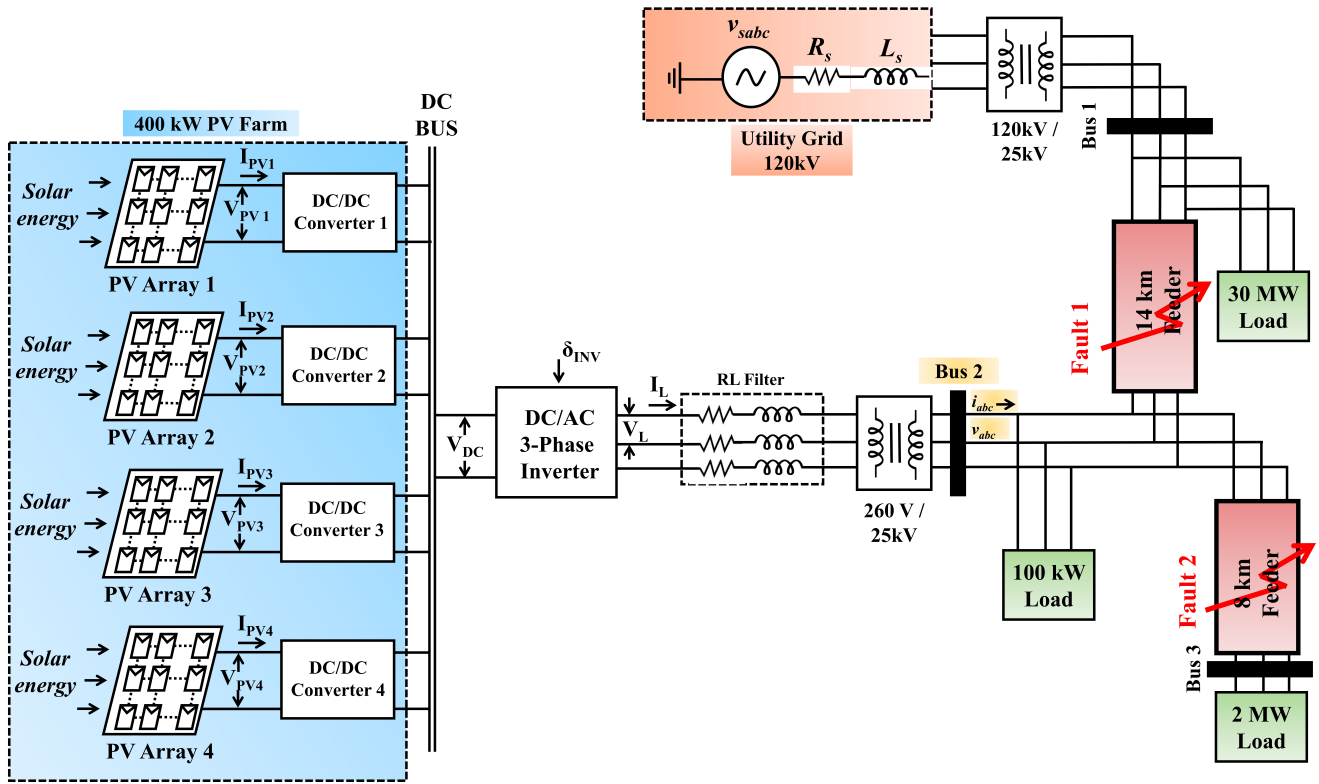}
  \caption{Schematic of the simulation model consisting of a 400 kW solar PV farm connected to the grid with three loads and two feeders.}
  \label{fig:schematic}
\end{figure*}

\subsection{Adaptation of SVM Kernel for Grassmann Manifold}
The Support Vector Machine (SVM) algorithm aims to establish an optimal hyperplane, effectively a decision boundary, to segregate an n-dimensional space into distinct classes. This facilitates the accurate categorization of new data points. In SVM, the extreme points or vectors crucial for forming this hyperplane are known as support vectors, defining the essence of the algorithm. A novel approach to apply kernel methods of SVM to manifold-valued data is discussed in \cite{Jayasumana2014KernelMO}. Kernel functions, particularly in SVM, serve to transform data into a high-dimensional space, simplifying the comparison of complex features.

Among various kernel functions, radial basis function (RBF) kernels are widely recognized for their general applicability and resemblance to the Gaussian distribution. The RBF kernel calculates the similarity between two points, say $Z_1$ and $Z_2$, in the following manner:
\begin{equation}
K(Z_{1},Z_{2}) = \exp\left(\frac{-||Z_{1}-Z_{2}||^{2}}{2\sigma^{2}}\right)
\end{equation}
Incorporating the Gaussian RBF kernel into the Grassmann manifold framework involves substituting the Euclidean distance with the distance metric defined earlier:
\begin{equation}
K_j([X_{1},X_{2}]) = \exp\left(-\beta , d_{j}^{2}([X_{1}],[X_{2}])\right)
\end{equation}
Here, $[X_i]$ represents the subspace spanned by the columns of $X_i$. Both $X_1$ and $X_2$ are matrices with orthonormal columns, and $\beta$ is a hyperparameter. This method applies the concept of mapping time series data to points on the Grassmann manifold for classification using SVM with the RBF kernel.

\section{Simulation Study, Result Analysis and Discussion}
\label{section:Simulation}

\subsection{Simulation Model}

The schematic of the simulation model is shown in Fig. \ref{fig:schematic}. The simulation is performed in MATLAB/SIMULINK. The considered system consists of a solar PV farm of four 100 kW PV arrays connected to a distribution system. The distribution system consists of 2 feeders of 8 km and 14 km supplying 3 loads. There are 3 buses in the system and the PV farm is connected to Bus 2, where the three-phase voltage and current measurements are measured and given as input to the proposed model for fault detection and clustering.

\begin{figure*}[]
\centering
\captionsetup[subfigure]{font=small}
\subfloat[$R_f = 0.01 \Omega$]{\includegraphics[width=0.24\textwidth]{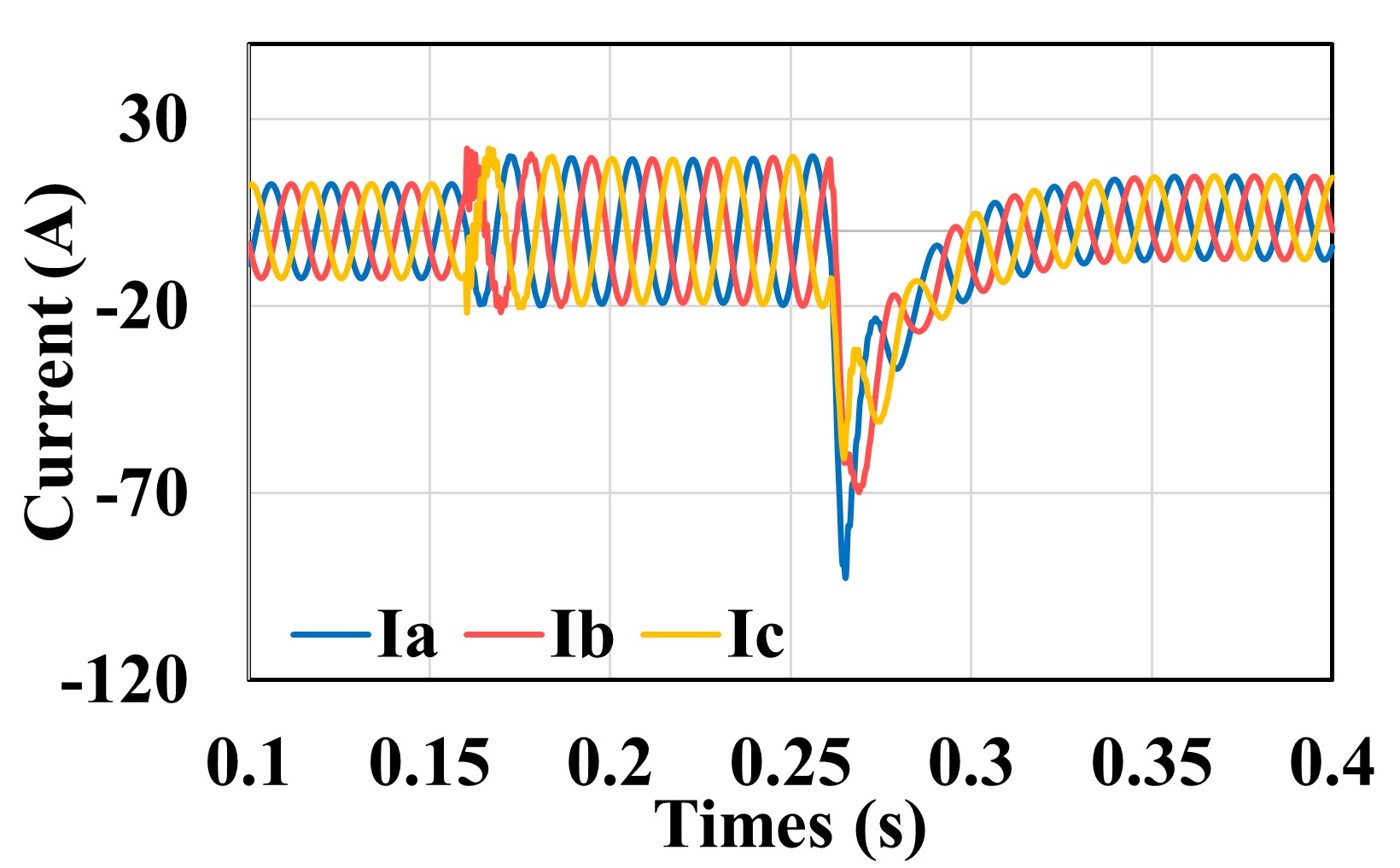}\label{fig:mode1}}\;
\subfloat[$R_f = 2 \Omega$]{\includegraphics[width=0.24\textwidth]{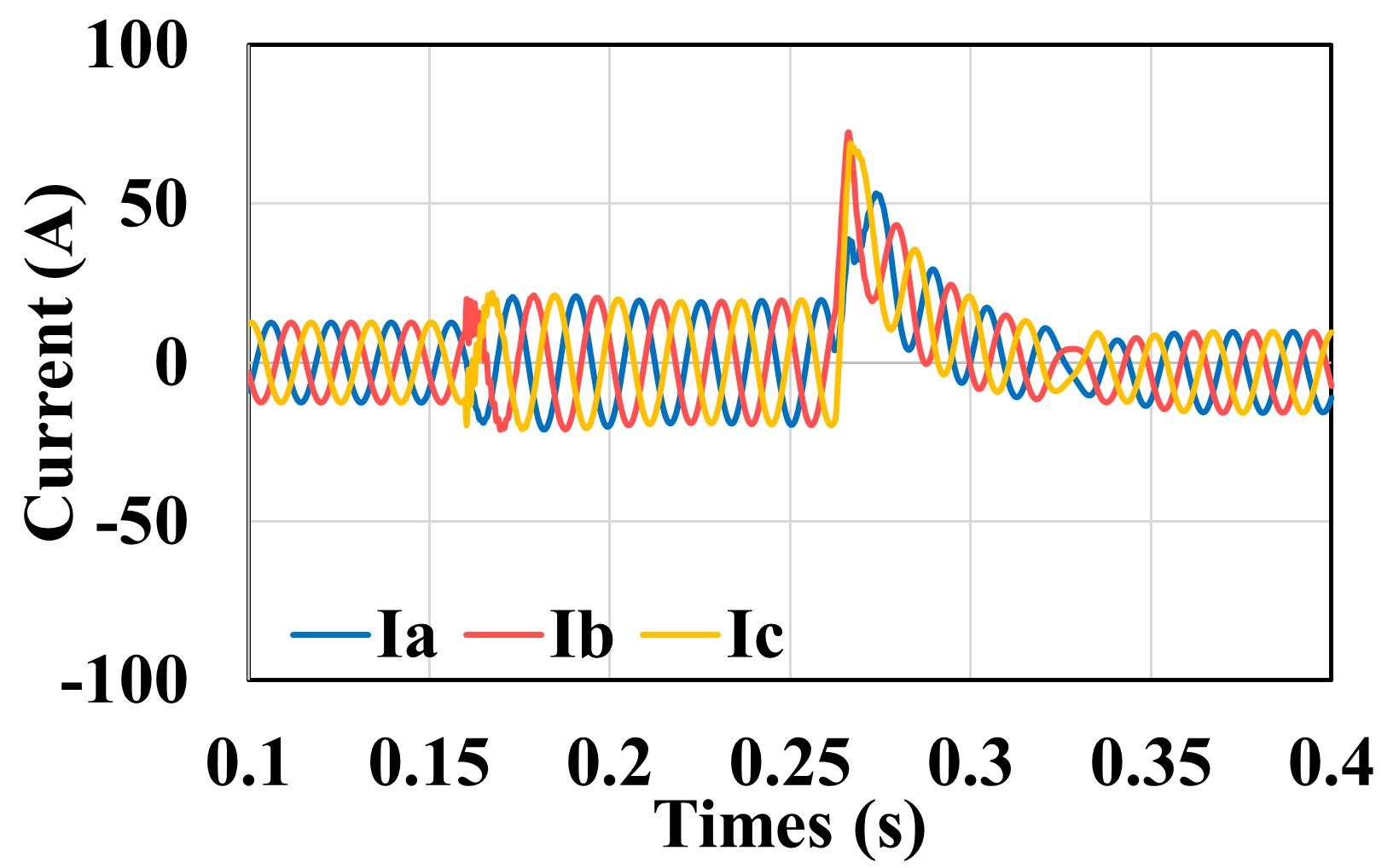}\label{fig:mode2}}\;
\subfloat[$R_f = 25 \Omega$]{\includegraphics[width=0.24\textwidth]{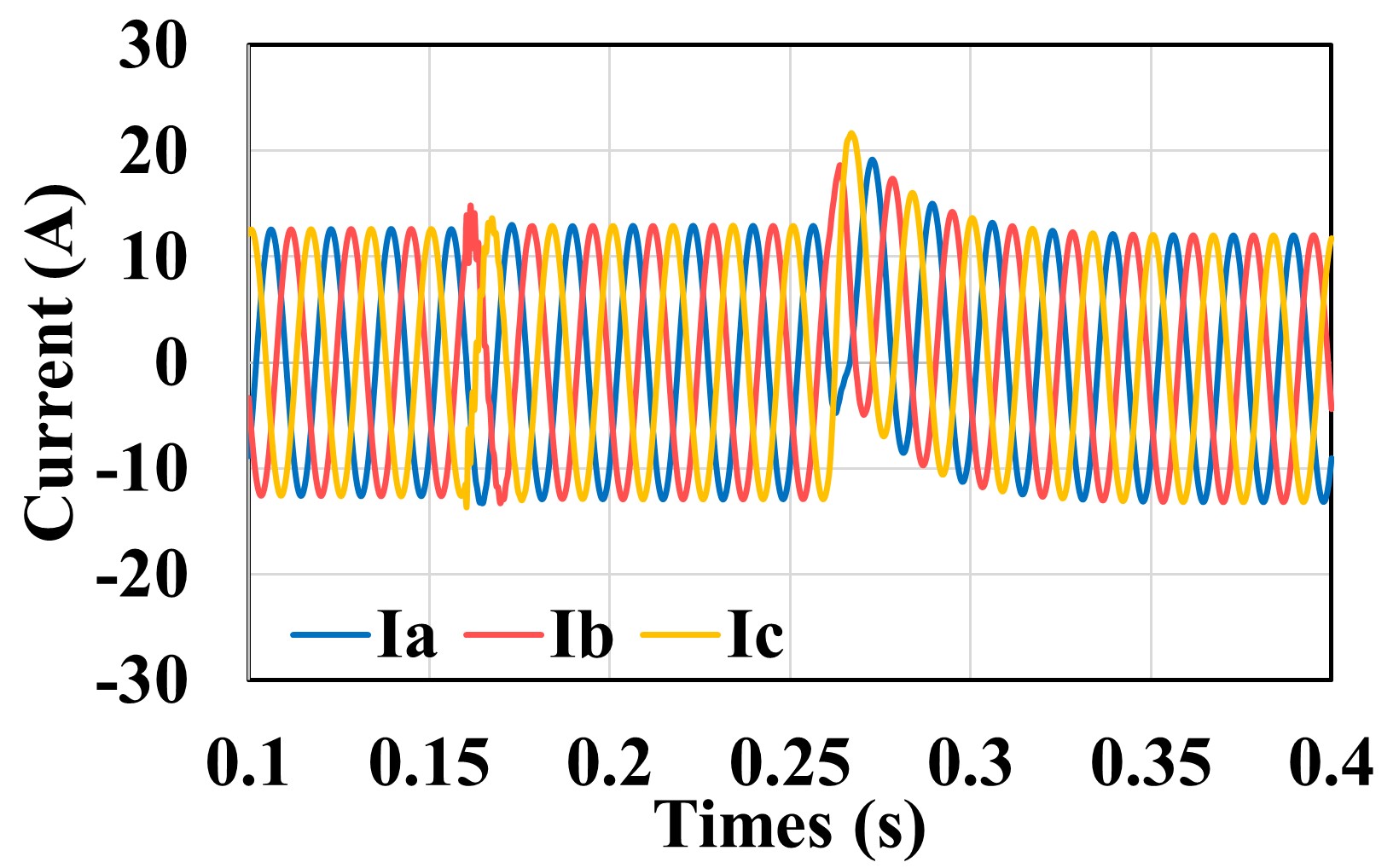}\label{fig:mode3}}\;
\subfloat[$R_f = 100 \Omega$]{\includegraphics[width=0.24\textwidth]{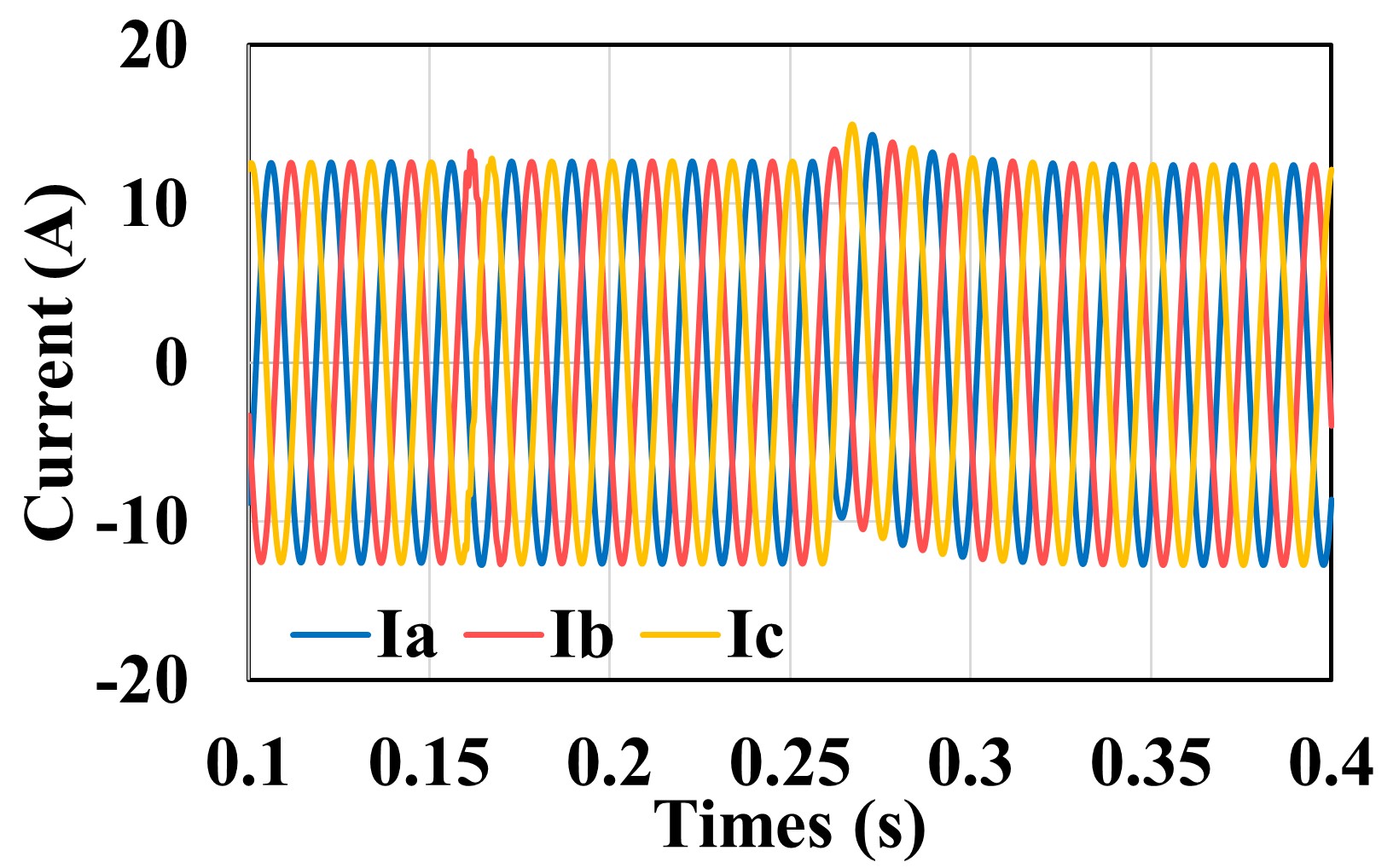}\label{fig:mode4}}\;
\caption{Fault with different fault resistances but same location and incident angle.}
\label{fig:diff_R}
\end{figure*}

\begin{figure*}
\centering
\captionsetup[subfigure]{font=small}
\subfloat[Line 1 - 1km]{\includegraphics[width=0.24\textwidth]{f1_b1_0_1.jpg}\label{fig:mode1}}\;
\subfloat[Line 1 - 10km]{\includegraphics[width=0.24\textwidth]{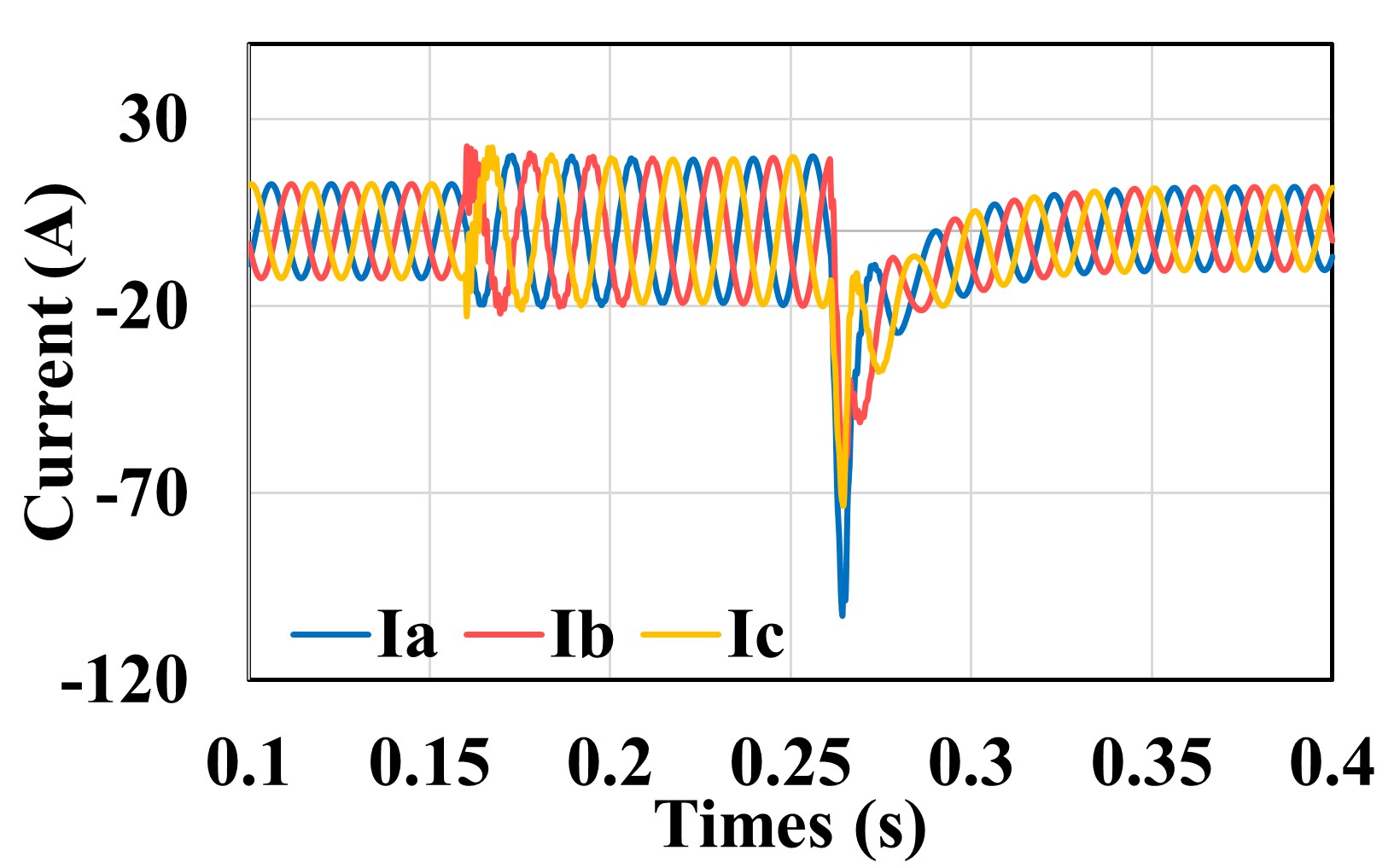}\label{fig:mode2}}\;
\subfloat[Line 2 - 1km]{\includegraphics[width=0.24\textwidth]{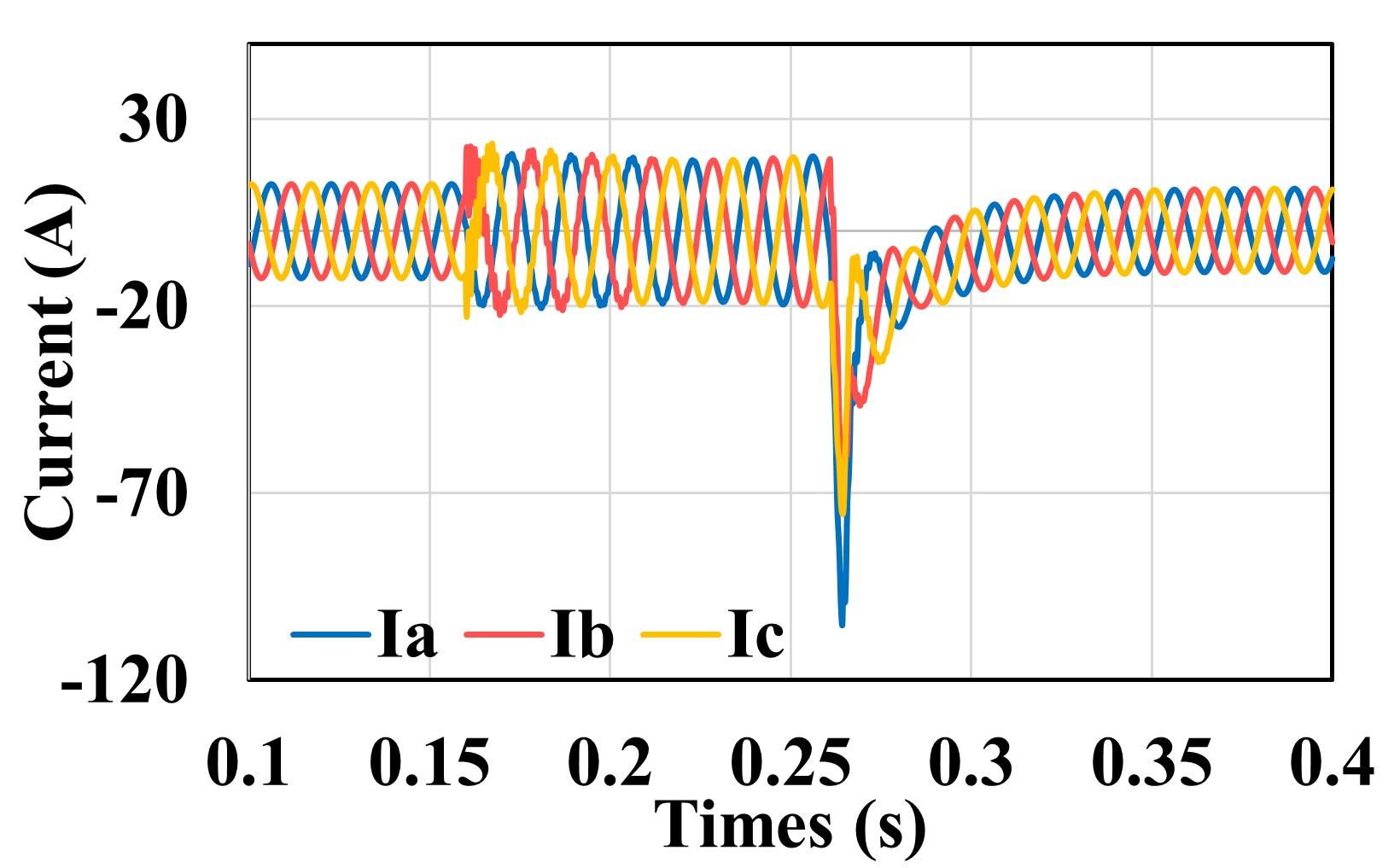}\label{fig:mode3}}\;
\subfloat[Line 2 - 8km]{\includegraphics[width=0.24\textwidth]{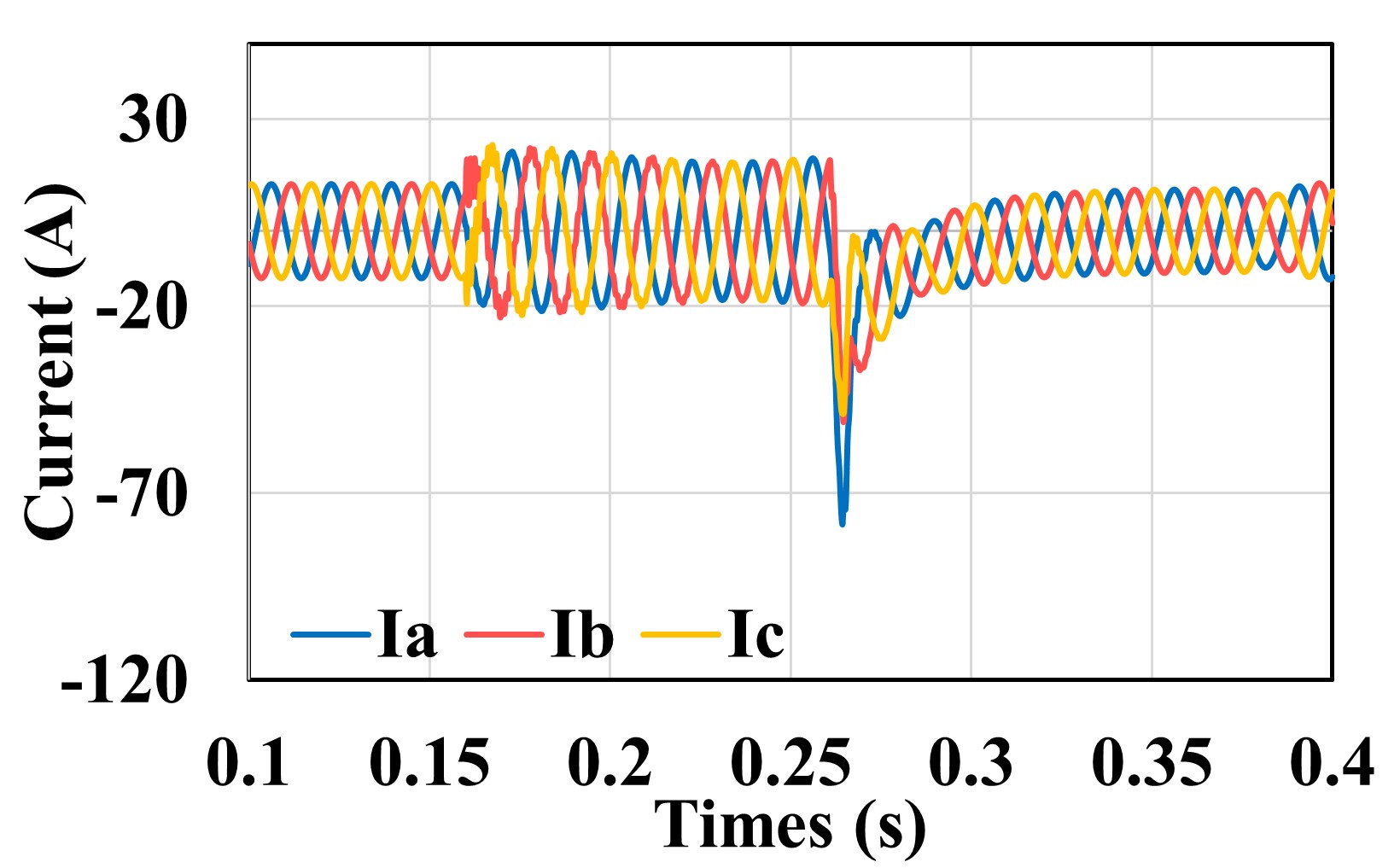}\label{fig:mode4}}\;
\caption{Fault with different fault locations but same resistances and incident angles.}
\label{fig:diff_locations}
\end{figure*}

\begin{figure*}
\centering
\captionsetup[subfigure]{font=small}
\subfloat[$\alpha = 0^0$]{\includegraphics[width=0.24\textwidth]{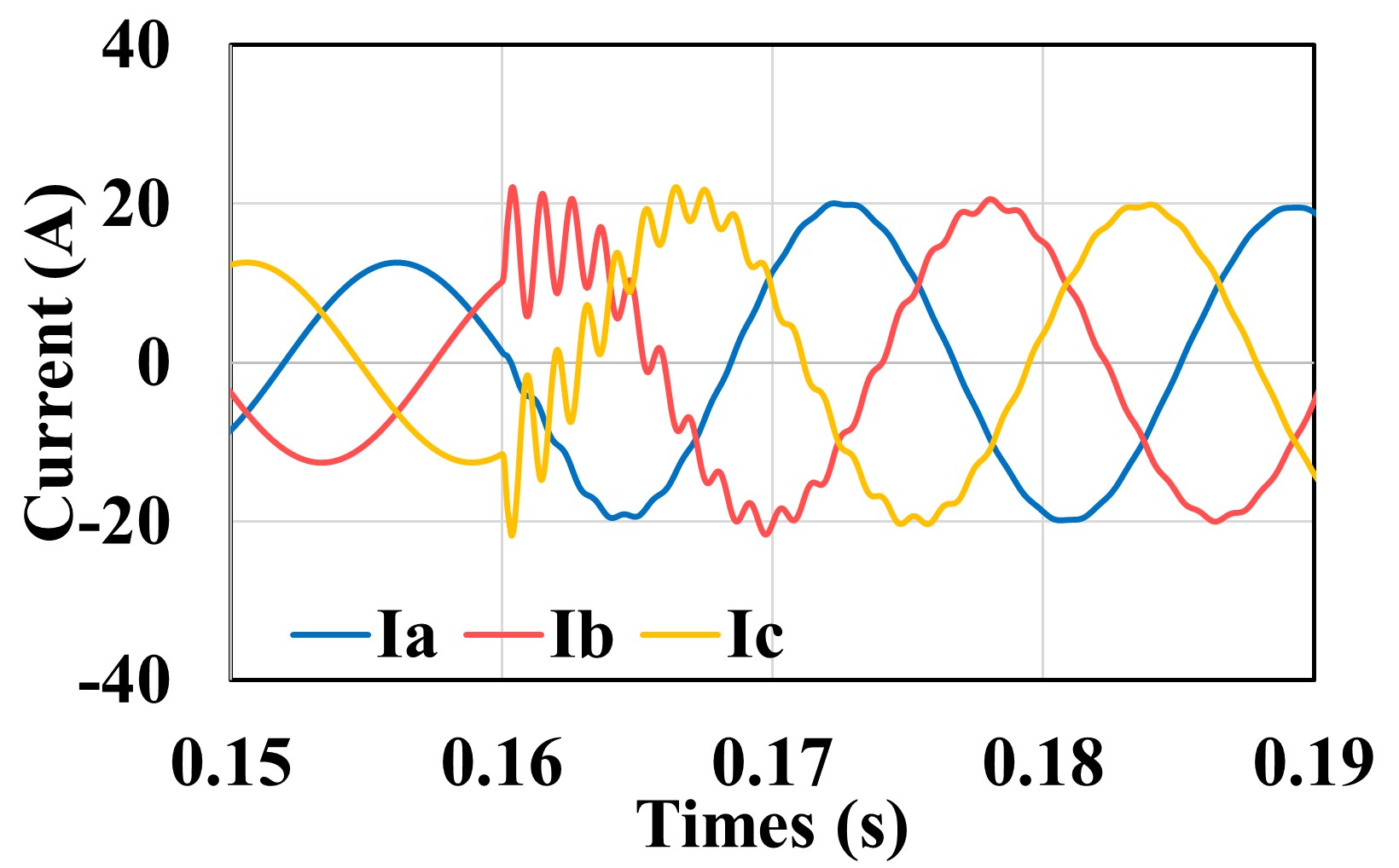}\label{fig:mode1}}\;
\subfloat[$\alpha = 60^0$]{\includegraphics[width=0.24\textwidth]{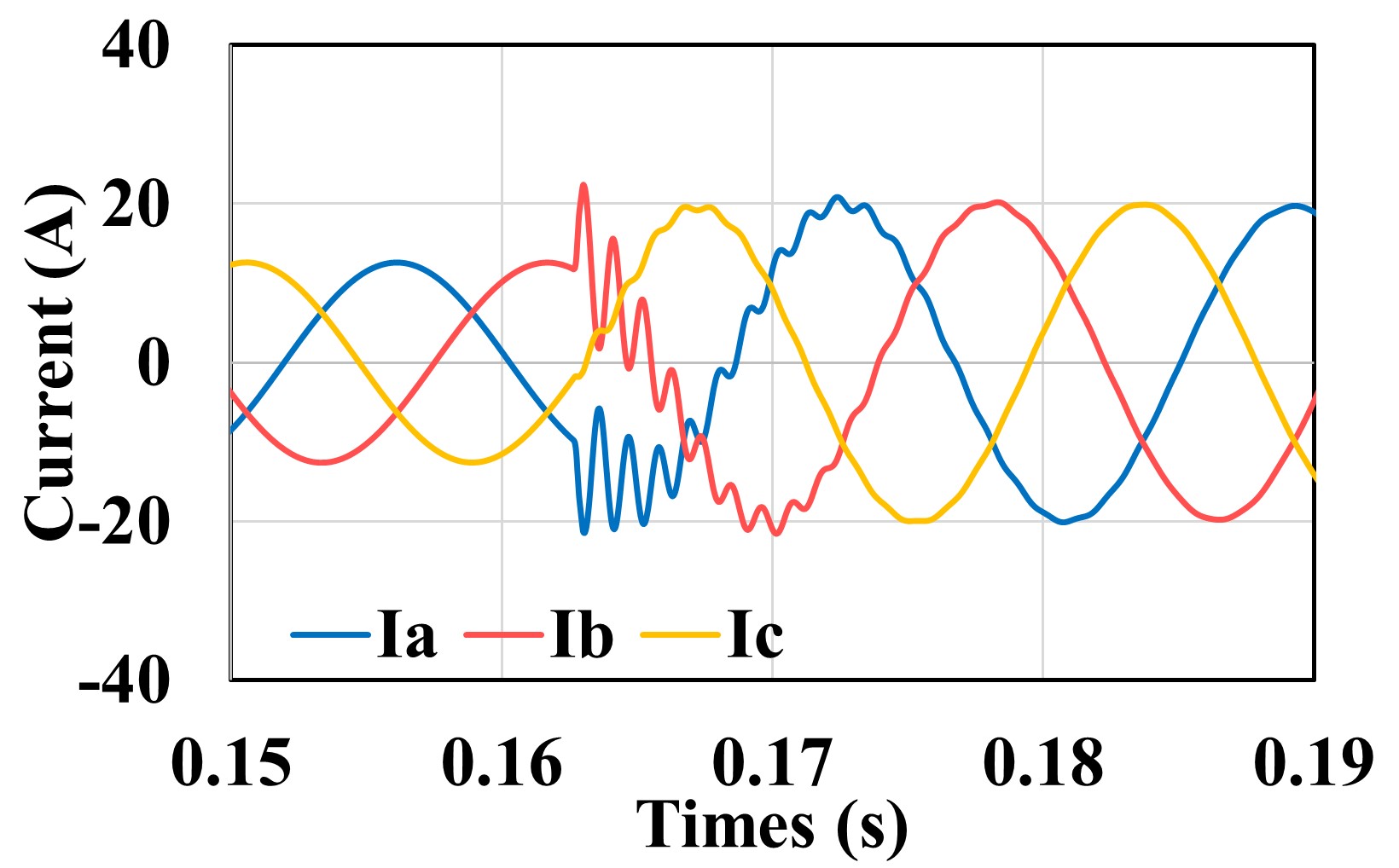}\label{fig:mode2}}\;
\subfloat[$\alpha = 120^0$]{\includegraphics[width=0.24\textwidth]{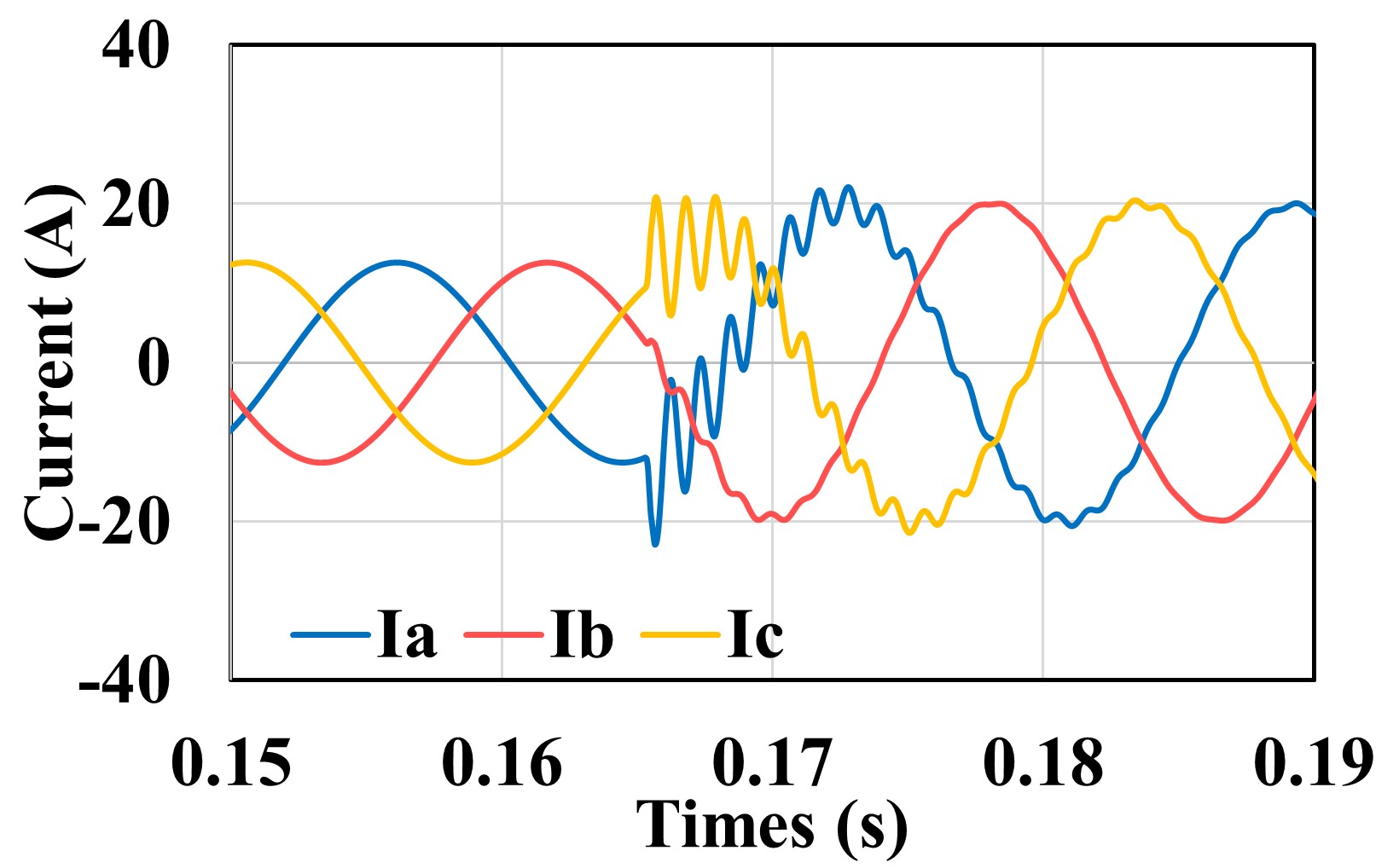}\label{fig:mode3}}\;
\subfloat[$\alpha = 180^0$]{\includegraphics[width=0.24\textwidth]{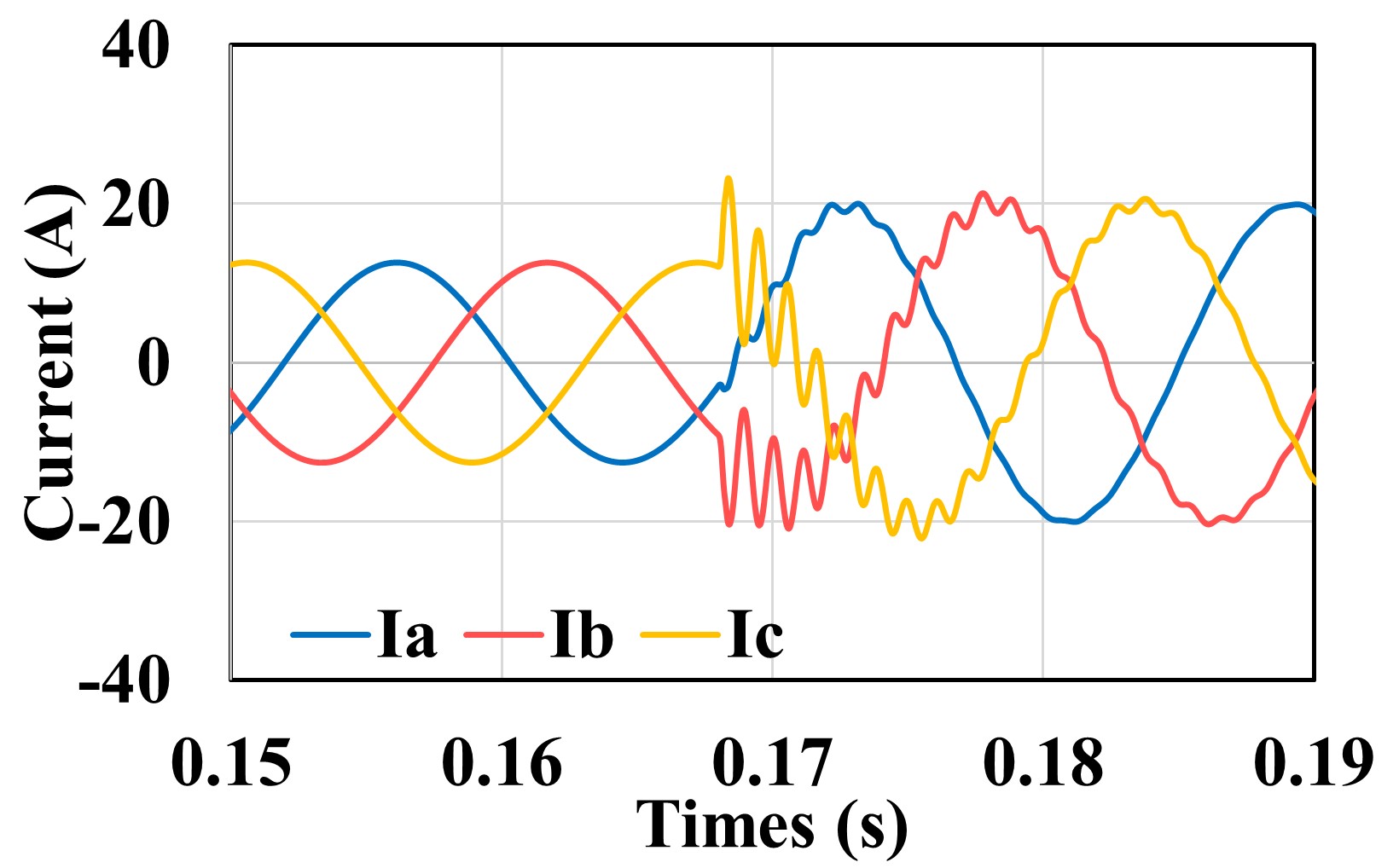}\label{fig:mode4}}\;
\caption{Fault with different fault incident angles but same resistances and locations.}
\label{fig:diff_angles}
\end{figure*}

The simulated fault cases are listed in Table \ref{tab:fault_cases}. Faults are considered on the two feeders for every 1 km distance; therefore, there are a total of 23 fault locations. Also, 9 different fault resistance values (0.01, 0.20, 2, 6, 10, 25, 50, 75, 100 ohms) are considered. Also, fault incident angles of 30\textdegree $\,$ variations are considered from 0\textdegree $\,$ to 180\textdegree. Thus, there are 1449 cases for each fault type. Eleven various types of fault are considered; single line-to-ground fault (LG), i.e., AG, BG, CG, double line-to-ground fault (LLG), i.e., ABG, BCG, CAG, double line fault (LL), i.e., AB, BC, CA, triple line-to-ground fault (TLG) i.e., ABCG and triple short-circuit fault (TSC) i.e., ABC. We have considered 2197 cases for no-fault conditions by varying the three loads from $\pm$ (0, 10, 20, 30, 40, 50, 60)\% of the rated load. The three-phase currents and voltages are measured at Bus 2 at the transformer connected to the PV farm. The current measurements for different fault resistances, different fault locations, and different fault incident angles are shown in Figs. \ref{fig:diff_R}, \ref{fig:diff_locations}, \ref{fig:diff_angles}, respectively. 

\begin{table}[b]
\centering
\caption{Fault Cases}
\begin{tabular}{cccccl}
 \toprule
   \begin{tabular}[c]{@{}c@{}}Case \\Type \end{tabular}&
   \begin{tabular}[c]{@{}c@{}}Type of \\Faults \end{tabular}&
    \begin{tabular}[c]{@{}c@{}}Fault \\Locations \end{tabular}&
  \begin{tabular}[c]{@{}c@{}}Fault \\Resistances \end{tabular}&
  \begin{tabular}[c]{@{}c@{}}Fault  \\Incident \\ Angles \end{tabular} \\
\midrule
Cases &
  \begin{tabular}[c]{@{}c@{}}LG Faults\\      LLG Faults\\      LL Faults\\      TSC \\    TLG \end{tabular} &
  \begin{tabular}[c]{@{}c@{}}Every 1km \\ of Feeder 1\\   and   Feeder 2\end{tabular} &
  \begin{tabular}[c]{@{}c@{}}0.01, 0.2, \\2.0, 6.0, \\      10, 25, \\50, 75, \\ 100\end{tabular} &
  \begin{tabular}[c]{@{}c@{}}0, 30,\\ 60, 90, \\      120, 150,\\ 180\end{tabular}
   \\
   \midrule
Total &
  11 &
  23 &
  9 &
  7 &\\
\bottomrule
\end{tabular}
\label{tab:fault_cases}
\end{table}

\begin{figure}[t]
  \centering
  \includegraphics[width=3.5in]{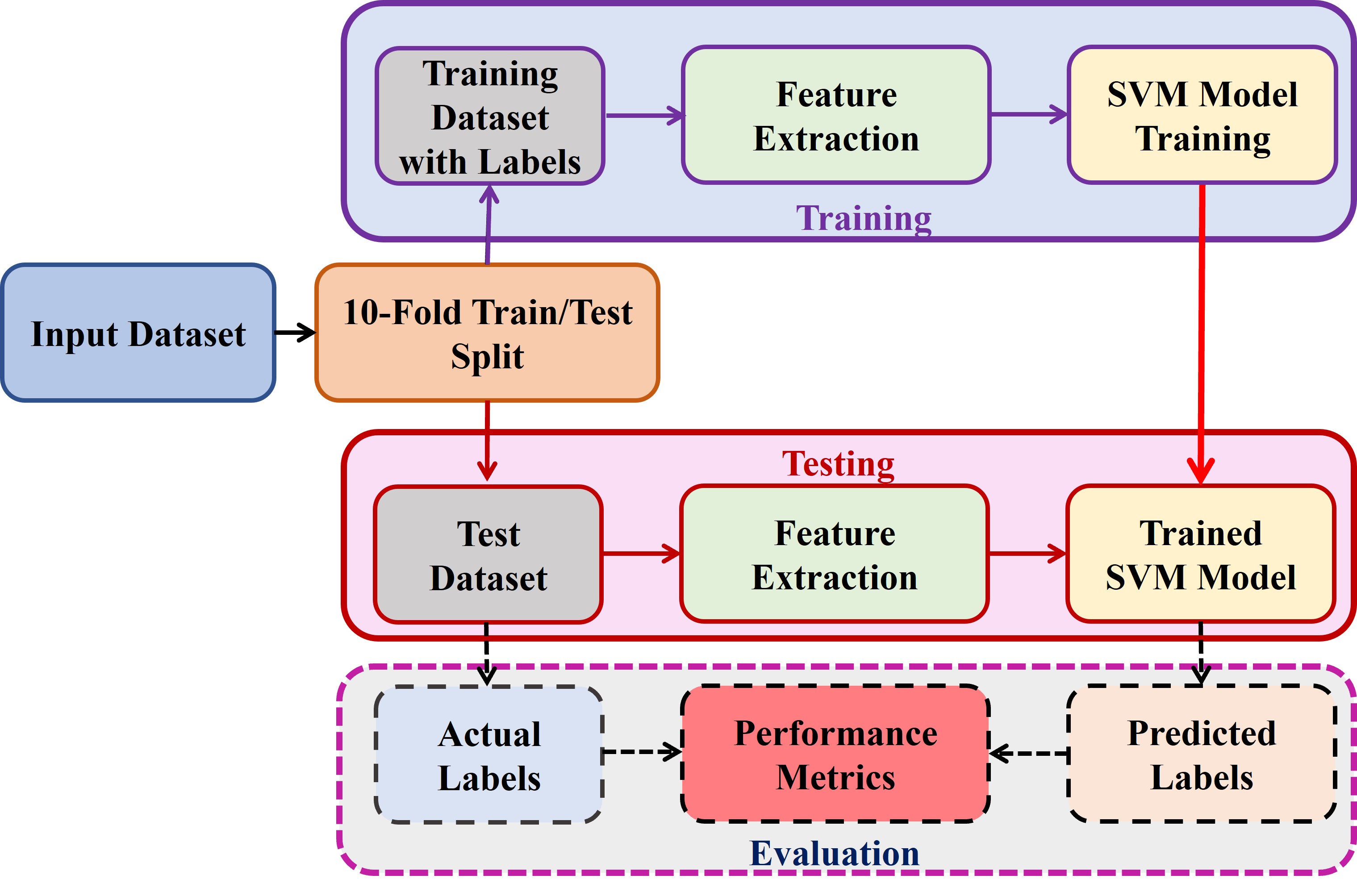}
  \caption{10-Fold cross validation.}
  \label{fig:10_fold_process}
\end{figure}

\subsection{Parameters for Grassmann Model Training}
The input dataset is a list of 2D arrays each representing a signal to be classified. For each element, the time samples are across axis 0 and features are along axis 1. For each sample we find its representation on the Grassmann manifold. This is done by first calculating the parameters of the ARMA model of the time series. After this the observability matrix is found and orthonormalized using these parameters which gives the Grassmann manifold representation. These extracted features are then passed to the SVM model with modified Gaussian kernel for classification.

In order to represent the input data on the Grassmann manifold, the recommended value of the number hidden dimensions range from 5-10 \cite{turaga2011statistical}. Since the number of hidden dimensions should be less than or equal to the number of features of the input data, here we have taken the hidden dimension to be 6. We have used the SVM model with Gaussian projection metric kernel for classification. In the context of SVM, the parameter $\gamma$ is a crucial hyperparameter that defines the width of the Gaussian. A smaller $\gamma$ leads to a smoother decision boundary and a larger value leads to a more complex and potentially tight-fitting decision boundary. The optimal value of $\gamma$ depends on the characteristics of the dataset and is found through experimentation. For our model, the value of $\gamma$ was set to 3. 
In order to evaluate the model generalization ability, we have performed 10 fold cross validation as shown in Fig. \ref{fig:10_fold_process} and in each fold the train set size was 90\% of the dataset and rest 10\% was included in the test set. In each fold, we extract features from the dataset by representing them on the Grassmann manifold. Then we train the SVM model with Gaussian kernel on the train set and evaluate the performance of the trained model on the test set using various performance metrics.

\subsection{Results and Discussion}


\begin{figure}
  \centering
  \includegraphics[width=3.4in]{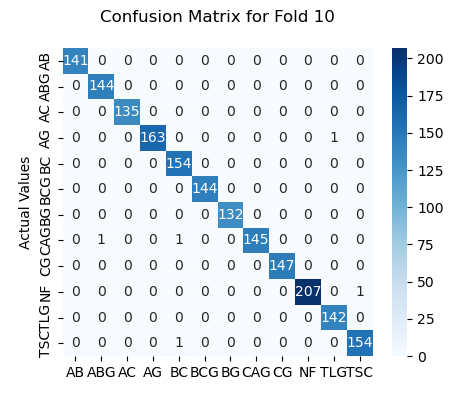}
  \caption{Confusion Matrix of Fold 10.}
  \label{fig:CM_fold10}
\end{figure}

The confusion matrix for Fold 10 of our model evaluation is illustrated in Fig. \ref{fig:CM_fold10}. Each row of the matrix represents the actual class labels, while each column corresponds to the predicted class labels. The diagonal elements indicate the number of instances where the predicted class matches the actual class, representing true positives for each respective class. For instance, class AB has 141 instances correctly identified, class AC has 144, and so forth. The off-diagonal elements represent the misclassifications made by the model; for example, class CG was mistaken once for class NF and once for class TLG. These misclassifications are minimal, underscoring the high predictive accuracy across classes. The robust performance of the classifier is evident from the concentration of high values along the diagonal, signifying a significant number of true positive predictions. Moreover, the sparse distribution of non-zero elements outside the diagonal suggests a low occurrence of false positives and false negatives, indicating the model's strong discriminative ability to correctly identify the classes. The evaluation metrics in Table \ref{tab:metrics} are calculated from the confusion matrix, and they provide further insight into the model's performance for each class.

\begin{table}[h!]
\centering
\caption{Evaluation metrics with formulae.}
\label{tab:metrics}
\begin{tabular}{lc}
\toprule
Metric & Formulae \\
\midrule
Accuracy & $\dfrac{TP + TN}{TP + TN + FP + FN}$ \\
Precision & $\dfrac{TP}{TP + FP}$ \\
Recall & $\dfrac{TP}{TP + FN}$ \\
MCC & $\dfrac{TP \times TN - FP \times FN}{\sqrt{(TP+FP) (TP+FN) (TN+FP) (TN+FN)}}$ \\
F1 - score & $2 \times \dfrac{(\text{Precision} \times \text{Recall})}{\text{Precision} + \text{Recall}}$ \\
\bottomrule
\end{tabular}

\vspace{1mm}

\raggedright{\quad TP -- True Positive, TN -- True Negative,\\ \quad FP -- False Positive, FN -- False Negative.}
\end{table}

\begin{table}[]
\centering
\caption{Evaluation metrics for each class.}
\begin{tabular}{cccccc}
\toprule
Class             & Accuracy & Precision & Recall  & F1-Score & Support \\
\midrule
AB                & 1        & 1         & 1       & 1          & 141     \\
ABG               & 0.99945  & 0.9931    & 1       & 0.99654    & 144     \\
AC                & 1        & 1         & 1       & 1          & 135     \\
AG                & 0.99945  & 1         & 0.9939  & 0.99694    & 164     \\
BC                & 0.9989   & 0.98718   & 1       & 0.99355    & 154     \\
BCG               & 1        & 1         & 1       & 1          & 144     \\
BG                & 1        & 1         & 1       & 1          & 132     \\
CAG               & 0.9989   & 1         & 0.98639 & 0.99315    & 147     \\
CG                & 1        & 1         & 1       & 1          & 147     \\
NF                & 0.99945  & 1         & 0.99519 & 0.99759    & 208     \\
TLG               & 0.99945  & 0.99301   & 1       & 0.99649    & 142     \\
TSC               & 0.9989   & 0.99355   & 0.99355 & 0.99355    & 155     \\
\midrule
M. Avg.    & 0.99954  & 0.99724   & 0.99742 & 0.99732    & 1813    \\
W. Avg. & 0.99952  & 0.99726   & 0.99724 & 0.99724    & 1813   \\
\bottomrule
\end{tabular}

\vspace{2mm}
\raggedright{ \quad M. Avg. - Macro Average, W. Avg. - Weighted Average}
\label{tab:Class_wise}
\end{table}

Table \ref{tab:Class_wise} presents the evaluation metrics for each class derived from the classification model's performance. The metrics included are accuracy, precision, recall, and F1-score, alongside the support (which represents the number of true instances for each class). The model demonstrates high precision and recall across most classes, as the scores are predominantly near to 1. Classes AB, BG, BCG, have scores of 1.00 across all metrics. The macro averages of the metrics are calculated as the arithmetic mean of the respective scores across all classes, thus treating each class equally irrespective of its support. These macro averages demonstrate the model's balanced capability across diverse classes, with values of 0.99594 for accuracy, 0.99724 for precision, 0.99742 for recall, and 0.99732 for the F1 score, indicating high overall performance. The weighted averages are computed by taking the support-weighted mean of the metrics, thereby accounting for the class imbalance by giving more weight to classes with higher support. The weighted averages similarly show high performance with scores of 0.99592 for accuracy, 0.99726 for precision, 0.99724 for recall, and 0.99724 for the F1-score.

\begin{table}[]
\centering
\caption{Weighted average of evaluation metrics for 10-fold cross-validation.}
\begin{tabular}{cccccc}
\toprule
Fold No. & Accuracy & Precision & Recall  & F1   Score & MCC     \\
\midrule
Fold 1    & 0.99947  & 0.99676   & 0.99669 & 0.99670    & 0.98313 \\
Fold 2    & 0.99975  & 0.99838   & 0.99835 & 0.99835    & 0.98916 \\
Fold 3    & 0.99942  & 0.99671   & 0.99669 & 0.99669    & 0.99397 \\
Fold 4    & 0.99981  & 0.99891   & 0.99890 & 0.99890    & 0.99096 \\
Fold 5    & 0.99991  & 0.99945   & 0.99945 & 0.99945    & 0.99819 \\
Fold 6    & 0.99966  & 0.99784   & 0.99779 & 0.99780    & 0.99217 \\
Fold 7    & 0.99965  & 0.99784   & 0.99780 & 0.99780    & 0.99036 \\
Fold 8    & 0.99973  & 0.99836   & 0.99835 & 0.99834    & 0.99396 \\
Fold 9    & 0.99954  & 0.99737   & 0.99699 & 0.99715    & 0.97890 \\
Fold 10   & 0.99952  & 0.99726   & 0.99724 & 0.99724    & 0.99397 \\
\midrule
Average  & 0.99965  & 0.99789   & 0.99783 & 0.99784    & 0.99048 \\
\bottomrule
\label{tab:10_fold}
\end{tabular}
\end{table}

The weighted average of evaluation metrics obtained from a rigorous 10-fold cross-validation is given in Table \ref{tab:10_fold}. This is performed to ensure the robustness and generalizability of the classification model. There are consistently high values across all metrics for each fold which indicate the model's stability and reliability in various subsets of the data. The accuracy remains above 0.9994 across all folds, reflecting the model's overall correctness. Precision and recall are similarly high, suggesting that the model is both relevant and comprehensive in its predictions. The F1-scores, which are the harmonic means of precision and recall, exceed 0.9967 for each fold, describing the model's balanced performance in terms of both false positives and false negatives. The MCC, a reliable statistical measure that accounts for true and false positives and negatives, is consistently above 0.9813; this further attests to the model's high quality of classification, considering that an MCC score of +1 represents a perfect prediction, 0 for an average random prediction, and -1 for an inverse prediction. The average of all the 10-folds reports the mean of each metric across all folds, presenting an aggregate measure of performance. With an average accuracy of 0.99965, precision of 0.99789, recall of 0.99783, F1-score of 0.99784, and MCC of 0.99048, the model demonstrates very high performance.

\section{Conclusion}
\label{section:Conclusion}

In this paper, we introduce a novel approach for electrical fault classification leveraging the Grassmann manifold representation. The Grassmann manifold is a mathematical framework capturing the geometry of subspaces, serves as the foundation for our methodology. We conducted simulations with a total of 15,939 fault cases, encompassing variations in fault resistances, locations, and incident angles, alongside 2,197 no-fault cases involving load variations in an electrical distribution system integrated with a PV farm. The proposed method demonstrates very high accuracy in terms of the classification of multiple classes without the use of any neural network model for training. This physics-based approach made use of the unique properties of the Grassmann manifold and demonstrated significant performance in fault classification due to the computational efficiency. This method is vital for enhancing the reliability and safety of power systems by facilitating timely maintenance and reducing downtime. As power systems continue to evolve and incorporate more complex components, our approach's ability to capture intricate fault patterns positions it as a valuable tool in ensuring the integrity and stability of these systems. Further, research and development in this direction could potentially lead to the integration of the proposed approach into practical power system monitoring and maintenance strategies.

\bibliographystyle{IEEEtran}
\bibliography{main}

\end{document}